\documentclass[
aps,prd,
12pt,
nopreprintnumbers,
showpacs,
eqsecnum,
nofootinbib
]{revtex4-1}

\usepackage{graphicx}

\def\Tr{{\rm Tr\,}}
\def\tr{{\rm tr\,}}

\def\v0{{\bf 0}}

\begin{document}

\title{
Graph-theory induced gravity and strongly-degenerate fermions in
a self-consistent Einstein universe}
\author{Nahomi Kan}\email[]{kan@yamaguchi-jc.ac.jp}
\affiliation{
Yamaguchi Junior College,
Hofu-shi, Yamaguchi 747--1232, Japan}
\author{Koichiro Kobayashi}\email[]{m004wa@yamaguchi-u.ac.jp}
\author{Kiyoshi Shiraishi}\email[]{shiraish@yamaguchi-u.ac.jp}
\affiliation{
Yamaguchi University,
Yamaguchi-shi, Yamaguchi 753--8512, Japan}
\date{\today}

\begin{abstract}
We study UV-finite theory of induced gravity. We use 
scalar fields, Dirac fields and vector fields as matter
fields whose one-loop effects induce the gravitational action.
To obtain the mass spectrum which satisfies the UV-finiteness condition,
we use a graph-based construction of mass matrices.
The existence of a self-consistent static solution for an Einstein
universe is shown in the presence of degenerate fermions. 
\end{abstract}



\maketitle

\section{Introduction}
The quantum nature of gravity is not yet cleared in spite of endeavor of
many researchers. An old idea on this issue is that gravity emerges as
quantum effects of matter fields \cite{IG}.
Originally, in such an induced gravity scenario, the Newton constant is
naturally obtained from the one-loop calculation with a cutoff of the
Planck scale. In this case, the induced cosmological constant becomes
a huge amount if no special choice of the matter-field content is
considered. 

We will consider a calculable model for induced gravity in the present
paper. For this purpose, we first fix the choice of matter species to
cancel the UV divergences. Next we should consider the mass spectra of
the fields, which affect the finite contribution to the induced Newton
constant and the cosmological constant.

To obtain the suitable mass spectra, we use the method of dimensional
deconstruction \cite{Deconstruction} and its generalization \cite{KSJMP}.
In the generalization of the deconstruction
model based on a graph, the eigenvalues of the graph Laplacian and the
adjacent matrix gives the mass spectrum of the particle. Thus we can
easily control the induced quantities at one-loop level in such a
model \cite{KSPTP}. 

We also study self-consistent static solutions for a static Einstein
universe in a graph-based induced gravity. 
We have considered self-consistent Einstein universe at finite
temperature in \cite{KSPT2}. In the present paper, we use the calculation
method with the spectral density function of the graph and search
for the static solution supported by the degenerate pressure of the
fermion at zero temperature. 

The present paper is organized as follows.
In \S2, we will examine the UV-divergences in field theory with the
heat kernel method.  The way to construct suitable models using the
knowledge of the graph structure is shown in
\S3.  In \S4, divergences in the effective gravitational action are
regularized for a static Einstein space.
It is shown that the technique with the density function to evaluate the
effective action for an Einstein space in \S5.  In \S6,
strongly-degenerate fermions and a self-consistent solution in our
model is studied. We give a summary and future prospects in
the last section.

\section{UV-finiteness condition}
Induced gravity  has been studied by many
authors \cite{IG}.
In terms of the heat kernel method \cite{hk}, the one-loop effective
action can systematically be expressed as an integral form using
Schwinger's proper time.

The classical action for a free field can be written as a quadratic form
with a differential operator on the spacetime manifolds.
The operator trace ($\Tr$) can be evaluated by the standard way to
rewrite
\begin{equation}
\frac{1}{2}\Tr\ln H=-\frac{1}{2}\int_0^\infty
\frac{dt}{t}\Tr\left[e^{-tH}\right]\,,
\end{equation}
where $H$ is a Hessian operator which appears in the free-field action.
The heat-kernel expansion can be expressed as, in four-dimensional
spacetime,
\begin{equation}
\Tr\left[e^{-tH}\right]=\frac{1}{(4\pi t)^2}\int
d^4x\sqrt{|\det g_{\mu\nu}|}\left[\tr a_0+t\,\tr a_1+t^2\,\tr
a_2+o(t^3)\right]\,,
\end{equation}
where 
$g_{\mu\nu}$ denotes the spacetime metric
and $\tr$ means the trace over the spacetime indices. The Seeley-DeWitt
coefficients $a_p$ $(p=0, 1, 2,\dots)$ depend on the background fields and
the first few coefficients have been known for several types of wave
operators. The one-loop effective action for the background fields is
given by the collection of the contribution of various matter fields to
the heat-kernel coefficients.

It is straightforward to see where the UV divergences occur, which we are
interested in.  The UV divergences arise from the integration in the
vicinity of $t=0$. These divergences arise from the first few terms of the
heat-kernel expansion. 
If we manage to introduce a UV-cutoff scale $\Lambda$,
the lower bound of the integration on $t$ is replaced to
${1/\Lambda^2}$.
To seek the condition for cancellation of UV divergences from various
matter fields, we need only  to consider massless fields. In the present
paper, minimally-coupled scalar fields, spinor fields, and vector fields
are taken into consideration.

The first Seeley-DeWitt coefficient $a_0$, which is a constant value,
has been found for such fields.
The value for each mode is:
$a_0=1$ for a scalar mode, $a_0=2$ for a spinor field, and
$a_0=2$ for a massless vector.
Then the effective Lagrangian at one-loop level includes the following
cutoff-dependent term proportional to \cite{IG}
\begin{equation}
\frac{1}{64\pi^2}(N_0-2N_{1/2}+2N_1)\Lambda^{4}\,,
\label{4div}
\end{equation}
where $N_0$ is the number of minimal scalar degrees of freedom,
$N_{1/2}$ is the number of two-component fermion fields, and
$N_1$ is the number of massless vector fields.
Note that the spinor field contributes with a negative sign for its
fermionic nature.
The expression (\ref{4div}) corresponds to the
cosmological constant or dark energy, if we treat it as a
cutoff-regularized theory.

The less divergent term comes from the coefficient $a_1$.
The coefficient for each mode is: 
$a_1=R/6$ for a scalar mode,
$a_1=-R/6$ for a spinor field, 
$a_1=-2R/3$ for a massless vector field,
where $R$ is the scalar curvature of the spacetime.
Thus the coefficient $a_1$ 
leads to the induced Einstein-Hilbert term.
The effective Lagrangian at one-loop level includes the following
cutoff-dependent term proportional to~\cite{IG}
\begin{equation}
\frac{1}{192\pi^2}(N_0+N_{1/2}-4N_1)\Lambda^2 R\,.
\end{equation}

Now we find that, to cancel the quartic and quadratic divergent terms,
which diverge as
$\Lambda\rightarrow \infty$,
we should choose
\begin{equation}
N_0=2N\,,\quad N_{1/2}=2N\,,\quad N_1=N\,,
\label{cond}
\end{equation}
where $N=1, 2, 3,\dots$.

The value of the Seeley-DeWitt coefficients can be confirmed when we set a
specific background-space geometry. Since the eigenvalues of the wave
operators for various fields on
$S^3$ are well known, 
the trace part
$\{{\rm tr}\,\exp\left[-Ht\right]\}$ for each field can be
evaluated as follows and has an asymptotic form for small
$t$: \cite{KSPT2} 
\begin{eqnarray}
\sum_{\ell=0}^\infty
(\ell+1)^2\exp\left[-\frac{\ell(\ell+2)}{a^2}t\right]
&=&\frac{2\pi^2a^3}{(4\pi
t)^{3/2}}\left(1+\frac{1}{a^2}t+\cdots\right), ~{\rm
for~a~scalar~mode}\,,\qquad\qquad
\label{s1}
\\
\sum_{\ell=1}^\infty
2\ell(\ell+1)\exp\left[-\frac{(\ell+1/2)^2}{a^2}t\right]
&=&\frac{2(2\pi^2a^3)}{(4\pi
t)^{3/2}}\left(1-\frac{1}{2a^2}t+\cdots\right), ~{\rm
for~a~spinor~field}\,,
\label{s2}
\\
\sum_{\ell=2}^\infty
2(\ell^2-1)\exp\left[-\frac{\ell^2}{a^2}t\right]
&=&\frac{2(2\pi^2a^3)}{(4\pi
t)^{3/2}}\left(1-\frac{2}{a^2}t+\cdots\right), ~{\rm
for~a~massless~vector}\,,
\label{s3}
\end{eqnarray}
where $a$ is the radius of $S^3$.
Because we know that the volume of $S^3$ is $2\pi^2 a^3$, the scalar
curvature of
$S^3$ is $6/a^2$ and the second-order (Euclidean) time-derivative
contribution gives a factor $(4\pi t)^{-1/2}$, the values of
$a_0$ and
$a_1$ for these fields mentioned above can be verified by
(\ref{s1}-\ref{s3}).

If the massless matter content satisfies the condition (\ref{cond}),
there is no quartic nor quadratic divergence and also no induced
gravitational action because of absence of mass scales.
Thus we should consider massses of the fields to yield the finite
contribution of quantum effects. Nonetheless, for cancellation of UV
divergences, the condition (\ref{cond}) is still necessary.

The algorithm to include the masses is very easy in the Schwinger time
integration. We only attach the following to the integrand for each
field
\begin{equation}
\sum_{i=1}^{N_s}e^{-(m^2_s)_i
t}=N_s-t\,\sum_{i=1}^{N_s}(m^2_s)_i+t^2\,\frac{1}{2}\sum_{i=1}^{N_s}
(m^4_s)_i+\cdots\equiv
N_s-t\,\Tr M^2_s+t^2\,\frac{1}{2}\Tr M^4_s+\cdots\,,
\end{equation}
where $M^2_s$ is the mass-squared matrix for spin-$s$ field.

In addition we need some interpretations in this trick.
For massive spinor fields, we replace $N'_{1/2}$ spinor fields to massive
$N'_{1/2}/2$ Dirac fields. For massive vector fields, we replace $N'_1$
massless vector fields as transverse modes and $N'_1$ scalar modes as
longitudinal modes to $N'_1$ massive vector fields. Now we find the
additional quadratic divergence is proportional to
\begin{equation}
\Tr M^2_S-4\,\Tr M^2_{D}+3\,\Tr M^2_V\,,
\end{equation}
where $M_S^2$ is the mass-squared matrix of $N'_0$ massive scalar fields,
$M_D^2$ is that of $N'_{1/2}/2$ massive Dirac fields, and
$M_V^2$ is that of 
$N'_1$ massive vector fields.

Finally, the condition for cancellation of the quartic and quadratic
divergences is concluded as follows.
The matter content is: $2N-N'_0-N'_1$ massless scalar fields, 
$2N-N'_{1/2}$ massless Weyl spinor fields, $N-N'_1$ massless vector
fields, $N'_0$ massive scalar fields, $N'_{1/2}/2$ massive Dirac fields,
and $N'_1$ massive vector fields.
Moreover, massive fields must have mass matrices which satisfy $\Tr
M^2_S-4\,\Tr M^2_{D}+3\,\Tr M^2_V=0$.

\section{Graph-based construction of a specific mass matrix}
In this section we construct the field theory with suitable mass
matrices which satisfy the UV-finite condition expressed in the previous
section.

Now we remember
the concept of dimensional deconstruction \cite{Deconstruction},
which is equivalent to considering a higher-dimensional theory with
discretized extra dimensions at a low-energy scale.
A moose diagram is used to describe
this theory, and is no more than a graph. 
The $N$-sided polygon is identified as an example of simple graphs, a
cycle graph $C_N$. 

A graph $G$ consists of a vertex set ${\cal V}$
and  an edge set ${\cal E}$,
where an edge is a pair of distinct vertices of $G$.
The degree of a vertex $v$, denoted by $deg(v)$, 
is the number of edges incident with $v$.
If all the degrees of vertices of a graph are equal, we  call such a
graph as a regular graph. 

We can consider the orientation of an edge. The
graph with directed edges is dubbed as a directed graph. An oriented edge
$e=[u,v]$  connects the 
origin $u=o(e)$ and the terminus $v=t(e)$.

Spectral graph theory is the mathematical study of a graph by
investigating various properties on eigenvalues, and eigenvectors of matrices
associated with it \cite{Mohar}.
Now we introduce various matrices that are naturally associated with a
graph \cite{KSJMP,Mohar} for later use.
   
The incidence matrix $E(G)$ is defined as 
\begin{equation}
(E)_{ve} = \left\{\begin{array}{rl}
            1 &  \text{if $v=o(e)$ }           \\ 
            -1 &  \text{if $v=t(e)$ }          \\ 
            0 &  \text{~otherwise}
         \end{array}\right. \, .
\end{equation}
The adjacency matrix $A(G)$ is defined as
\begin{equation}
(A)_{vv^{\prime}} =\left\{\begin{array}{rl}
            1 &  \text{if $v$ is adjacent to $v^{\prime}$ }   \\ 
            0 &  \text{otherwise}
          \end{array}\right. .
\end{equation}
The degree matrix $D(G)$ is defined as
\begin{equation}
(D)_{vv^{\prime}} =\left\{\begin{array}{rl}
           deg(v) &  \text{if $v = v^{\prime}$ }           \\ 
            ~~ 0   &  \text{otherwise}
        \end{array}\right. .
\end{equation}
Note that ${\rm Tr}\,A=0$ and ${\rm Tr}\,A^2={\rm Tr}\,D$,
and for a regular graph, $D$ is proportional to the identity matrix.

The graph Laplacian (or combinatorial Laplacian) 
$\Delta (G)$ is defined as
\begin{equation}
(\Delta)_{vv^{\prime}} 
      = (D-A)_{vv^{\prime}}
      = \left\{\begin{array}{rl}
            deg(v) &  \text{if $v = v^{\prime}$ }           \\ 
             -1    &  \text{if $v$ is adjacent to $v^{\prime}$}        \\
            ~~ 0   &  \text{otherwise}
          \end{array}\right. .
\end{equation} 
The most important observation is    
\begin{equation}
\Delta= EE^T\,,
\label{EE}
\end{equation} 
where $E^T$ is the transposed matrix of $E$.
The Laplacian matrix is symmetric, so its eigenvalues are non-negative.
Note also that ${\rm Tr}\,\Delta={\rm Tr}\,D$ and
${\rm Tr}\,\Delta^2={\rm Tr}\,D^2+{\rm Tr}\,D$.

The simplest model of vector fields has been studied by Hill and
Leibovich \cite{HL}. The generalized model associated with a general
graph is written down as~\cite{KSJMP}
\begin{equation}
{\cal L}_V=-\frac{1}{4}\sum_{v\in {\cal V}}F^v_{\mu\nu}F_v^{\mu\nu}-
\sum_{e\in {\cal E}}({\cal D}_\mu U_e)^\dagger ({\cal D}^\mu U_e)\,,
\end{equation} 
where the covariant derivative is
\begin{equation}
{\cal D}^\mu U_e\equiv (\partial^\mu+iA^\mu_{t(e)}-iA^\mu_{o(e)})U_e\,,
\end{equation} 
with $|U_e|=f$, $f$ is a constant with the dimension of mass.
The vector fields $A^\mu_v$ are assigned at vertices of $G$ and the scalar
fields
$U_e$ are assigned at edges of $G$ in this model.

Similarly, any kind of fields can be associated with a graph
and their mass-squared matrix can be written using
the graph Laplacian.
For scalar fields, we assign a scalar field
$\phi_v$ to each
vertex $v$ of
$G$. A difference can be defined on each edge
$e$ as
\begin{equation}
d\phi_e\equiv \phi_{t(e)}-\phi_{o(e)}=-\sum_{v\in {\cal
V}}E^T_{ev}\phi_v\,.
\end{equation}
Thus a mass term for scalar fields can be constructed as
\begin{equation}
f^2\sum_{e\in {\cal E}}d\phi_e d\phi_e=f^2\sum_{e\in {\cal
E}}\sum_{v,v'\in {\cal V}}\phi_{v'}E_{v'e} E^T_{ev}
\phi_v=f^2\sum_{v,v'\in {\cal V}}\phi_v\Delta_{vv'}
\phi_{v'}\,.
\end{equation}

For spinor fields, the mass term can be expressed using
the incidence matrix
$E$. 
For example, the Lagrangian density of fermion fields can be written
as \cite{KSJMP}
\begin{equation}
-\sum_{v\in {\cal V}}\bar{\psi}_{Rv}{\it
D\!\!\!\!/~}\psi_{Rv}-\sum_{e\in {\cal E}}\bar{\psi}_{Le}{\it
D\!\!\!\!/~}\psi_{Le}-f\sum_{e\in
{\cal E}}\sum_{v\in {\cal V}}[(\bar{\psi}_{Le}(E^T)_{ev}\psi_{Rv}+h.
c.]\,,
\label{fermionlag}
\end{equation}
where the subscripts $L$ and $R$ denote left-handed and right-handed
fermions, respectively.
Namely, the left-handed fermions are assigned to the edges while the right-handed
ones are assigned to the vertices. The mass-squared matrix for $\psi_{Rv}$
is expressed as
$f^2EE^T=f^2\Delta$ while that for $\psi_{Le}$ is $f^2E^TE\equiv
f^2\tilde\Delta$. The matrices $\Delta$ and
$\tilde\Delta$ have the same spectrum up to zero modes. Thus the mass spectrum of
fermions governed by the Lagrangian (\ref{fermionlag}) is also given by
the eigenvalues of the graph Laplacian (\ref{EE}). For details, see
Ref.~\cite{KSJMP}.
                          
With the knowledge in spectral graph theory \cite{Mohar},
we can find that the UV divergent terms are 
concerned with the graph Laplacian.
Therefore, the UV divergences can be controlled by using the graph
Laplacian and we can construct the models of UV-finite induced gravity
from spectral graph theory.

A prescription is as follows.
First we prepare three graphs, $G_S$, $G_D$ and $G_V$.
All these graphs have $N$ vertices.
We can construct Lagrangians whose mass-squared matrices satisfy
\begin{equation}
\Tr M_S^2=\Tr M_D^2=\Tr M_V^2\,,\quad
\Tr M_S^4=\Tr M_D^4=\Tr M_V^4\,,
\end{equation}
by choosing graphs as $D(G_S)=D(G_D)=D(G_V)$ \cite{KSPTP}.
Then we find that the induced vacuum energy at one-loop level is
\cite{KSPTP}
\begin{equation}
V_0=-\frac{1}{(4\pi)^2}\int_0^\infty
\frac{dt}{t^3}\,\Tr\left[e^{-M_S^2t}-4e^{-M_D^2t}+3e^{-M_V^2t}\right]\,,
\end{equation}
and the inverse of the Newton constant is given by \cite{KSPTP}
\begin{equation}
\frac{1}{16\pi {\rm G}}=-\frac{1}{6(4\pi)^2}\int_0^\infty
\frac{dt}{t^2}\,\Tr\left[e^{-M_S^2t}+2e^{-M_D^2t}-3e^{-M_V^2t}\right]\,.
\end{equation}
In the flat-space limit, the one-loop vacuum energy has been calculated
for field theory associated with the cycle graph $C_n$ \cite{KSPTP}. The
degree matrix of a cycle graph $C_n$ is an $n\times n$ diagonal matrix
$diag.(2,2,\dots,2)$. We select a type of non-simply-connected graphs
$G_{\{n_i\}}=C_{n_1}\cup C_{n_2}\cup
\cdots=\bigcup_{\sum_i n_i=N}C_{n_i}$, which has $N$ vertices. The degree
matrix of
$G_{\{n_i\}}$ is an $N\times N$ diagonal matrix $diag.(2,2,\dots,2)$.
Therefore, if the mass-squared matrix $M^2$ is proportional to the graph
Laplacian of
$G_{\{n_i\}}$, $\Tr M^2$ and $\Tr (M^2)^2$ are independent of the choice
of the set $\{n_i\}$, as long as $\sum_i n_i=N$ is fixed. 
We can choose different sets $\{n_i\}$ for scalar, Dirac, and vector field
model in order to obtain non-zero value for the Newton and cosmological
constants \cite{KSPTP}.

\section{evaluation of the effective action in $S^3$ with zeta functions}
We will consider a model for
the static universe with spatial topology
$S^3$ with the radius $a$, in later sections.
The self-consistent induced gravity model at finite temperature has been
studied in Ref.~\cite{KSPT2}. 
We will study degenerate fermions at zero
temperature and the self-consistent universe later in the present paper.

In this section, we evaluate the one-loop vacuum energy for the spacetime
$R\times S^3$. To this end, we use (\ref{s1}-\ref{s3}) in the Schwinger
integral form of the effective action. Here we first integrate over the
proper-time
$t$, but then we slightly shifted the power of $t$ in the integrand. For
example, an expression which appears in the effective action is 
rewritten as
\begin{eqnarray}
& &\int_0^{\infty}\frac{dt}{t^{3/2-s}}\sum_{\ell=0}^\infty
(\ell+1)^2\exp\left[-\frac{\ell(\ell+2)}{a^2}\,t-m^2 t\right]\nonumber \\
&=&\frac{\Gamma(s-1/2)}{a^{1-2s}}\sum_{\ell=0}^\infty
\frac{(\ell+1)^2}{\left[\ell(\ell+2)+m^2a^2\right]^{s-1/2}} 
\,.
\end{eqnarray}
We then look for where divergences occur.
We follow an analogous method used in Ref.~\cite{nash}, to separate a
convergent summation from others.
Now we convert it to
\begin{eqnarray}
&&\Sigma_S(m^2a^2)\equiv\sum_{\ell=0}^\infty
\frac{(\ell+1)^2}{\left[\ell(\ell+2)+m^2a^2\right]^{s-1/2}}=\sum_{\ell=1}^\infty
\frac{\ell^2}{\left[\ell^2+m^2a^2-1\right]^{s-1/2}}\nonumber \\
&=&\sum_{\ell=1}^\infty\left[
\frac{\ell^2}{\left[\ell^2+m^2a^2-1\right]^{s-1/2}}-
\frac{1}{\ell^{2s-3}}\left(1+\frac{(1/2-s)(m^2a^2-1)}{\ell^2}+
\frac{(s^2-1/4)(m^2a^2-1)^2}{2\ell^4}\right)\right]\nonumber \\
&
&+\zeta_R(2s-3)+(1/2-s)(m^2a^2-1)\zeta_R(2s-1)+
\frac{(s^2-1/4)(m^2a^2-1)^2}{2}\zeta_R(2s+1)\,,
\end{eqnarray}
where $\zeta_R(z)$ is the Riemann's zeta function.
Similarly we find
\begin{eqnarray}
\Sigma_D(m^2a^2)&\equiv&4\sum_{\ell=1}^\infty
\frac{\ell(\ell+1)}{\left[(\ell+1/2)^2+m^2a^2\right]^{s-1/2}}=
4\sum_{\ell=0}^\infty
\frac{(\ell+1/2)^2-1/4}{\left[(\ell+1/2)^2+m^2a^2\right]^{s-1/2}}\nonumber
\\ &=&4\sum_{\ell=0}^\infty\left[
\frac{(\ell+1/2)^2-1/4}{\left[(\ell+1/2)^2+m^2a^2\right]^{s-1/2}}\right.\nonumber
\\ &&\qquad -\left.
\frac{(\ell+1/2)^2-1/4}{(\ell+1/2)^{2s-1}}\left(1+
\frac{(1/2-s)m^2a^2}{(\ell+1/2)^2}+
\frac{(s^2-1/4)m^4a^4}{2(\ell+1/2)^4}\right)\right]\nonumber \\
&&+4\left\{(2^{2s-3}-1)\zeta_R(2s-3)+\left[\left(\frac{1}{2}-s\right)m^2a^2-
\frac{1}{4}\right]
(2^{2s-1}-1)\zeta_R(2s-1)\right.\nonumber \\
&&\qquad+\left.\left[
\frac{(s^2-1/4)m^4a^4}{2}-\frac{(1/2-s)m^2a^2}{4}\right](2^{2s+1}-1)
\zeta_R(2s+1)\right.\nonumber \\
&&\qquad-\left.
\frac{(s^2-1/4)m^4a^4}{2}(2^{2s+3}-1)\zeta_R(2s+3)\right\}\,,
\end{eqnarray}
and also
\begin{eqnarray}
\Sigma_V(m^2a^2)&\equiv&2\sum_{\ell=2}^\infty
\frac{\ell^2-1}{\left[\ell^2+m^2a^2\right]^{s-1/2}}=2\sum_{\ell=1}^\infty
\frac{\ell^2-1}{\left[\ell^2+m^2a^2\right]^{s-1/2}}\nonumber \\
&=&2\sum_{\ell=1}^\infty\left[
\frac{\ell^2-1}{\left[\ell^2+m^2a^2\right]^{s-1/2}}-
\frac{\ell^2-1}{\ell^{2s-1}}\left(1+\frac{(1/2-s)m^2a^2}{\ell^2}+
\frac{(s^2-1/4)m^4a^4}{2\ell^4}\right)\right]\nonumber \\
&&+2\left\{\zeta_R(2s-3)+\left[\left(\frac{1}{2}-s\right)m^2a^2-1\right]
\zeta_R(2s-1)\right.\nonumber \\
&&+\left.\left[
\frac{(s^2-1/4)m^4a^4}{2}-\left(\frac{1}{2}-s\right)m^2a^2\right]\zeta_R(2s+1)-
\frac{(s^2-1/4)m^4a^4}{2}\zeta_R(2s+3)\right\}\,.\nonumber \\
\end{eqnarray}
Since $\zeta_R(-3)=\frac{1}{120}$
and
$\zeta_R(-1)=-\frac{1}{12}$ are finite,
only divergent part for $s\rightarrow 0$ in each $\Sigma$ is the term
including $\zeta_R(2s+1)$. The divergent parts are
\begin{equation}
2\Sigma_S^{div}(m^2a^2)=
\left(s^2-\frac{1}{4}\right)(m^2a^2-1)^2\zeta_R(2s+1)\,,
\end{equation}
\begin{equation}
\Sigma_D^{div}(m^2a^2)=
\left[
\left(2s^2-\frac{1}{2}\right)m^4a^4-\left(\frac{1}{2}-s\right)m^2a^2\right](2^{2s+1}-1)
\zeta_R(2s+1)\,,
\end{equation}
\begin{equation}
\Sigma_V^{div}(m^2a^2)
=
\left[
\left(s^2-\frac{1}{4}\right)m^4a^4-\left(1-2s\right)m^2a^2\right]
\zeta_R(2s+1)\,.
\end{equation}

In the graph-based model reviewed in the previous section,
we can set $\Tr M_S^4=\Tr M_D^4=\Tr M_V^4$ as well as
$\Tr M_S^2=\Tr M_D^2=\Tr M_V^2$.
Thus the divergence in the induced action is proportional to
\begin{equation}
\lim_{s\rightarrow
0}\sum_i[2\Sigma^{div}_S(m^2_ia^2)-\Sigma^{div}_D(m^2_ia^2)+
\Sigma^{div}_V(m^2_ia^2)]=N\lim_{s\rightarrow 
0}\left(-\frac{1}{4}+O(s)\right)\zeta_R(2s+1)\,.
\end{equation}
This residual divergence is only in $\Sigma_S$ and independent of mass, in
other words, it appears even in the case with massless (minimal) scalar
fields. Elizalde \cite{Elizalde} argued that this divergence should be
dealt by `principal part prescription'.
In the prescription, the pole term in the Riemann's zeta function is
discarded.
This minimal subtraction yields
\begin{equation}
\zeta_R(2s+1)=\frac{1}{2s}+\gamma+O(s)\rightarrow \gamma\,,
\end{equation}
where $\gamma$ is the Euler-Mascheroni constant ($\gamma\approx
0.577216$).

Apart from the divergence,
the divergent terms up to $m^4$ have been canceled.
Corresponding to the analysis by using integral form \`ala Schwinger,
we find that divergences including mass parameter can be cancelled in our
graph-based models.

We now redefine the finite part of summations as
\begin{eqnarray}
\Sigma'_S(m^2a^2)&=&\sum_{\ell=1}^\infty \ell^2\left[
\sqrt{\ell^2+m^2a^2-1}-
\ell \left(1+\frac{m^2a^2-1}{2\ell^2}-
\frac{(m^2a^2-1)^2}{8\ell^4}\right)\right]\nonumber \\
&
&+\zeta_R(-3)+\frac{m^2a^2-1}{2}\zeta_R(-1)-\frac{1}{8}\gamma\,,
\end{eqnarray}
\begin{eqnarray}
\Sigma'_D(m^2a^2)&=&4\sum_{\ell=0}^\infty
\left[(\ell+1/2)^2-1/4\right]\left[
\sqrt{(\ell+1/2)^2+m^2a^2}\right.\nonumber
\\ &&\qquad -\left.
(\ell+1/2)\left(1+
\frac{m^2a^2}{2(\ell+1/2)^2}-
\frac{m^4a^4}{8(\ell+1/2)^4}\right)\right]\nonumber \\
&&+4\left\{-\frac{7}{8}\zeta_R(-3)-\left[\frac{1}{4}m^2a^2-
\frac{1}{8}\right]
\zeta_R(-1)+
\frac{7m^4a^4}{8}\zeta_R(3)\right\}\,,
\end{eqnarray}
and
\begin{eqnarray}
\Sigma'_V(m^2a^2)
&=&2\sum_{\ell=1}^\infty(\ell^2-1)\left[
\sqrt{\ell^2+m^2a^2}-
\ell\left(1+\frac{m^2a^2}{2\ell^2}-
\frac{m^4a^4}{8\ell^4}\right)\right]\nonumber \\
&&+2\left\{\zeta_R(-3)+\left[\frac{1}{2}m^2a^2-1\right]
\zeta_R(-1)+
\frac{m^4a^4}{8}\zeta_R(3)\right\}\,.
\end{eqnarray}

Then we find the effective action in the form,
\begin{equation}
\frac{1}{2a}
\sum_i\left[\Sigma'_S((m_0^2)_ia^2)-
\Sigma'_D((m_{1/2}^2)_ia^2)+\Sigma'_V((m_{1}^2)_ia^2)\right]\,.
\end{equation}

\section{use of spectral density function of a graph}
In this section, we introduce the spectral density function of a
graph \cite{HO}.
The use of the spectral density makes the analysis of the
Casimir energy very easy.
In the present paper, we consider only regular graphs.
Remembering that the graph Laplacian is expressed as $\Delta=D-A$,
we need only to consider the spectral density function for the adjacency
matrix
$A$ in the case with a regular graph.

We start with the case for a cycle graph $C_N$, for example.
The spectrum of the eigenvalues for the adjacency matrix of $C_N$
is
\begin{equation}
\lambda_k=2\cos\frac{2\pi k}{N}\,,\quad (k=0, 1, \dots, N-1)
\end{equation}
and thus the eigenvalues for $\Delta$ are $\Lambda_k=2-2\cos\frac{2\pi
k}{N}=4\sin^2\frac{\pi k}{N}$.
It has been shown \cite{HO} that, since
\begin{equation}
\lim_{N\rightarrow\infty}\frac{1}{N}
\sum_{k=0}^{N-1}f\left(\lambda_k\right)=
\int_0^1f(2\cos \pi
t)dt=\frac{1}{\pi}\int_{-2}^{2}f(x)\frac{dx}{\sqrt{4-x^2}}\,,
\end{equation}
the spectral density in the large $N$ limit can be employed as
\begin{equation}
\lim_{N\rightarrow\infty}\int_{-\infty}^{+\infty}f(x)
\rho_N(x) dx=\frac{1}{\pi}\int_{-2}^{2}f(x)\frac{dx}{\sqrt{4-x^2}}\,.
\end{equation}
Namely, the summation about the discrete eigenvalues becomes an
integration over the continuous variable
$x$ with the spectral density function 
$\rho_\infty(x)$,
\begin{equation}
\rho_\infty(x)=\left\{
\begin{array}{cc}
\frac{1}{\pi}\frac{1}{\sqrt{4-x^2}}\, &\quad {\rm for~}-2<x<2\\
0\, &\quad {\rm otherwise}
\end{array}
\right.\qquad {\rm for~cycle~graphs}\,,
\end{equation}
in the large $N$ limit.
Incidentally, the precise spectral density function for $C_N$
with a finite $N$ is known as
\begin{equation}
\rho(x)=\left\{
\begin{array}{cc}
\frac{1}{\pi}\frac{1}{\sqrt{4-x^2}}[1+2\sum_{k=1}^{\infty}
T_{kN}(x/2)]\,
&\quad {\rm for~}-2<x<2\\ 0\, &\quad {\rm otherwise}
\end{array}
\right.\,,
\end{equation}
where $T_n(z)$ denotes the Chebyshev polynomial.

The spectral density function is known for other several graphs.
The trace formula for regular graph $G$ of degree $q+1$ on $N$ vertices
is  \cite{tf}
\begin{equation}
\frac{1}{N}\sum_{i=1}^{N}e^{t\lambda_i}=\frac{q+1}{2\pi}
\int_{-2\sqrt{q}}^{2\sqrt{q}}e^{xt}\frac{\sqrt{4q-x^2}}{(q+1)^2-x^2}dx
+\frac{1}{N}\sum_g\sum_{k=1}^\infty
\frac{\ell(g)}{2^{k\ell(g)/2}}
I_{k\ell(g)}(2\sqrt{q}t)\,,
\end{equation}
where $g$ runs over the set of all oriented primitive closed
geodesics in $G$, and $\ell(g)$ is the length of $g$, while $I_n(z)$
is the modified Bessel function of the first kind.
Then
\begin{equation}
\rho_\infty(x)=\left\{
\begin{array}{cc}
\frac{q+1}{2\pi}
\frac{\sqrt{4q-x^2}}{(q+1)^2-x^2}\,
&\quad {\rm for~}-2\sqrt{q}<x<2\sqrt{q}\\ 0\, & {\rm otherwise}
\end{array}
\right.\qquad {\rm for~}(q+1){\rm -regular~graphs}\,.
\end{equation}

In the present paper, we will concentrate ourselves on the case with the
graph
$G_{\{n_i\}}=C_{n_1}\cup C_{n_2}\cup
\cdots=\bigcup_{\sum_i n_i=N}C_{n_i}$.
Clearly enough, one find that the spectral density function $\rho_\infty$
is independent of the choice of $\{n_i\}$.

This fact implies that the finite contributions for the Newton and
cosmological constant come from the $\rho_N-\rho_\infty$
if the summation is evaluated as the integration over the continuum
variables.
Therefore the Casimir energy behaves as $1/a^4\times 2\pi^2 a^3$ and the
similar contribution which dominates if $a$ is small are
substantially calculated only by using $\rho_\infty$
and that is independent of values for the Newton and cosmological constant
in the flat-space limit.
This universal conclusion may be interesting if we try to extend the
present approach to the case with general graphs.

Turning to the present analysis, we assume that the mass-squared matrix is
given by
$f^2\Delta(G)$, where
$f$ is a unique mass scale in the model.
For large $N$, the effective action, where the Casimir energy is dominant,
becomes
\begin{equation}
\Omega_0(fa)\equiv\frac{1}{2a}\int_{-2}^2\left[\Sigma'_S(f^2a^2(2-x))-
\Sigma'_D(f^2a^2(2-x))+\Sigma'_V(f^2a^2(2-x))\right]
\frac{N}{\pi\sqrt{4-x^2}}dx\,.
\end{equation}

In the next section, using this result, we study a self-consistent
cosmological solution for an Einstein universe in the graph-based
induced gravity model. 

\section{degenerate fermions and a self-consistent universe} 
We consider a model for
the static universe with spatial topology
$S^3$ with the radius $a$.
The static homogeneous, closed space is often called an Einstein
universe. 
The self-consistent induced gravity model at finite temperature has been
studied in Ref.~\cite{KSPT2}. 
In the present paper, we study the self-consistent cold universe at 
zero temperature and we will consider degenerate fermions.
Although the cold universe seems to have less relevance to the actual
universe than the hot case, it can be a possible phase between quantum
cosmology and classical cosmology. 

In the static spacetime, it is known that the effective action can be
interpreted as the total free energy of the quantum fields at
finite temperature \cite{KKEU}. Similarly, we consider the thermodynamic
potential for the case with a finite chemical potential.

The thermodynamic
potential of a system of strongly-degenerate fermionic fields at zero
temperature can be computed as \cite{cd}
\begin{equation}
\Omega_D=-\frac{2\pi^2 a^3}{12\pi^2}\sum_i\theta(\mu-m_i)
\left[\mu\sqrt{\mu^2-m_i^2}\left(\mu^2-\frac{5}{2}m_i^2\right)+
\frac{3}{2}m_i^4\ln\left(\frac{\mu}{m_i}+
\sqrt{\frac{\mu^2}{m_i^2}-1}\right)\right]\,,
\end{equation}
where $\mu$ is the chemical potential and $\theta(y)$ is the step
function,
$\theta(y)=1$ for
$y\ge 0$ and 
$\theta(y)=0$ for $y<0$.

For the case with the model associated with the
graph which consists of a set of $C_n$, $\Omega_D$ in the large $N$
(the total number of vertices)
limit can be reduced to
\begin{eqnarray}
\Omega_D&=&-\frac{2\pi^2
a^3}{12\pi^2}\int_{-2}^2\theta(\mu-m(x))\nonumber \\ &\times&
\left[\mu\sqrt{\mu^2-m^2(x)}\left(\mu^2-\frac{5}{2}m^2(x)\right)+
\frac{3}{2}m^4(x)\ln\left(\frac{\mu}{m(x)}+
\sqrt{\frac{\mu^2}{m^2(x)}-1}\right)\right]\nonumber
\\&\times&\frac{N}{\pi\sqrt{4-x^2}} dx\,,
\end{eqnarray}
with $m^2(x)\equiv f^2(2-x)$.

It is known that the fastest way to obtain self-consistent equations is
by using the total free energy in the finite-temperature
case \cite{EU1}. Similarly, we consider the total thermodynamic
potential $\Omega$ as the sum of the contribution of quantum effects
$\Omega_0$ derived in the previous section and that of degenerate Dirac
fields $\Omega_D$.
 The energy of
the system is given by
\begin{equation}
E=\Omega+\mu{\cal
N}=\frac{\partial(\mu^{-1}\Omega)}{\partial(\mu^{-1})}\,,
\end{equation}
where
\begin{equation}
{\cal N}=-\frac{\partial\Omega}{\partial\mu}\,,
\end{equation}
is the fermion number, which suffers no correction from $\Omega_0$.
The  pressure $P$ is obtained by 
\begin{equation}
P\times(2\pi^2a^3)=-\frac{1}{3}a\frac{\partial\Omega}{\partial a}\,,
\end{equation}
as in the finite-temperature case.

The self-consistent equations can be derived as  
\begin{equation}
\frac{\partial(\mu^{-1}\Omega)}{\partial(\mu^{-1})}=0\,,
\end{equation}
and 
\begin{equation}
\frac{\partial(\mu^{-1}\Omega)}{\partial a}=0\,,
\end{equation}
where the first equation corresponds to the $00$-component of the
Einstein equation with one-loop corrections
and the second corresponds to the diagonal component in a spatial direction.
Thus the extremal point of $\mu^{-1}\Omega(fa,f/\mu)$ provides a solution
to the self-consistent equation.

\begin{figure}[h]
\centering
\includegraphics
{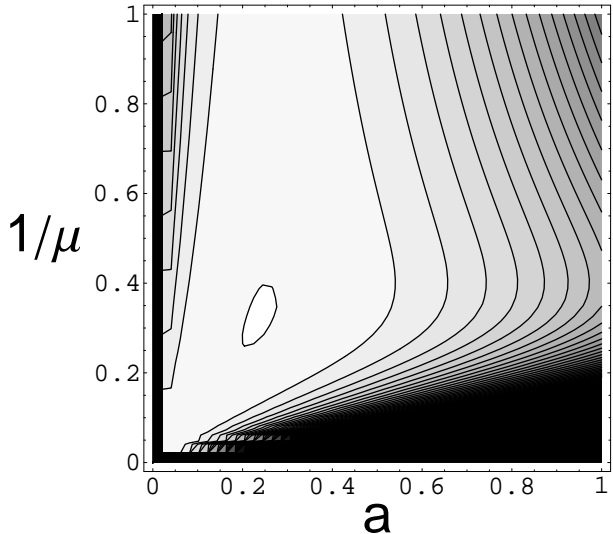}
\caption{%
A contour plot of $\frac{1}{N}\mu^{-1}\Omega$.
A solution of the self-consistent equation can be found at the maximum
point. 
}
\label{fig1}
\end{figure}

In FIG.~\ref{fig1}, we show the contour plots for $\Omega/\mu$ obtained by
numerical calculations, whose extremum provides a self-consistent
solution. The horizontal axis indicates the scale factor
$a$, while the vertical one $1/\mu$, in the unit of $f$.

Since 
the Casimir energy is dominant for small $a$,
the solution 
can be found at the maximum of $\mu^{-1} \Omega$, corresponding to
the Casimir regime defined in Ref.~\cite{EU1}.
The stability is not expected, for the extremum of the potential is
actually the maximum point.

\section{Summary and outlook}
In the present paper, we have examined ultra-violet divergences of a
one-loop calculable model for induced gravity.
We have found that finite values for the Newton and cosmological constant
can be realized if the mass-squared matrices for scalar, spinor, and
vector fields satisfy a few conditions.

It has been found that the model which has the suitable mass matrices can
be obtained by the graph-based construction.
In this paper, we focused on a type of the regular graph such as
$G=C_{n_1}\cup C_{n_2}\cup\cdots$.

To evaluate the effective action for an Einstein universe,
we need the knowledge of graph spectrum.
We have introduced the spectral density function of the graph and found
that it is useful to calculate the Casimir-energy dominant case, for
small $a$ and large $N$.

The spectral density is also convenient to evaluate the thermodynamical
potential of strongly-degenerate fermions.
We have studied self-consistent Einstein universe at zero temperature
with degenerate fermions in our model. We found that the Casimir regime
can been seen.

In the present analysis, we have constructed models using cycle graphs, 
but we are also interested in the model of general graphs.
As future works,
trace formula for a regular graph \cite{tf} will be useful.

The universal behavior of the
effective action for large $N$ and small $a$ under the condition of the
fixed type of the associated graph, is
interesting. 
If the construction of the model with dynamical selection of graphs is
possible, say, utilizing the Higgs-like mechanism assigned at edges or
vertices, it can be imagined that many large-scale universe with different
Newton and cosmological constants would develop once from a single state
with a large Casimir energy.
Anyway, we should investigate some variation of the present model.

\begin{acknowledgments}
The authors would like to thank the organizers of JGRG21, where our
partial result 
was presented.
\end{acknowledgments}



\bibliographystyle{apsrev4-1}

\end{document}